\documentclass[final,1p, times]{elsarticle}

\usepackage{amssymb}
\usepackage{graphics}
\usepackage{graphicx}
\usepackage{dcolumn}
\usepackage{bm}
\usepackage{booktabs}
\usepackage{hyperref}
\usepackage{subfigure}
\usepackage{multirow}
\usepackage{float}

\journal{Nuclear Physics A}

\begin{document}

\begin{frontmatter}



\title{Universal scaling of the pion, kaon and proton $p_{\rm{T}}$ spectra in Pb-Pb collisions at 2.76 TeV}

\author{Yanyun Wang, Liwen Yang, Xiaoling Du, Na Liu, Liyun Qiao}
\author{Wenchao Zhang\corref{cor1}}
\cortext[cor1]{Corresponding author}
\ead{wenchao.zhang@snnu.edu.cn}
\address{School of Physics and Information Technology, Shaanxi Normal University, Xi'an 710119, P. R. China}

\begin{abstract}
  With the experimental data collected by the ALICE collaboration in Pb-Pb collisions at a center-of-mass energy per nucleon pair 2.76 TeV for six different centralities (0-5$\%$, 5-10$\%$, 10-20$\%$, 20-40$\%$, 40-60$\%$ and 60-80$\%$), we investigate the scaling property of the pion, kaon and  proton transverse momentum ($p_{\rm{T}}$) spectra at these centralities. We show that in the low $p_{\rm{T}}$ region with $p_{\rm T} \leq$ 2.75 (3.10 and 2.35) GeV/c the pion (kaon and proton) spectra exhibit a scaling behaviour independent of the centrality of the collisions.  This scaling behaviour arises when these spectra are presented in terms of a suitable variable, $z=p_{\rm{T}}/K$. The scaling parameter $K$ is determined by the quality factor method and is parameterized by $a \langle N_{\rm{part}}\rangle^{b}$, where $\langle N_{\rm{part}}\rangle$ is the average value of the number of participating nucleons, $a$ and $b$ are free parameters, $b$ characterizes the rate at which $\textrm{ln} K$ changes with $\textrm{ln} \langle N_{\rm{part}}\rangle$. The values of $b$ for pions and kaons are consistent within uncertainties, while they are smaller than that for protons. In the high $p_{\rm{T}}$ region, due to the suppression of the spectra, a violation of the proposed scaling is observed going from central to peripheral collisions. The more peripheral the collisions are, the more clearly violated the proposed scaling becomes. In the framework of the colour string percolation model, we argue that the pions, kaons and protons originate from the fragmentation of clusters which are formed by strings overlapping and the cluster's fragmentation functions are different for different hadrons.  The scaling behaviour of the pion, kaon and proton spectra in the low $p_{\rm T}$ region can be simultaneously explained by the colour string percolation model in a qualitative way.

\end{abstract}
 

\end{frontmatter}



\section{Introduction}\label{Intro}
The transverse momentum ($p_{\rm T}$) spectra of final state particles play an important role in understanding the dynamics of particle production in high energy collisions. Several approaches are utilized to search for the dynamics, one of them being the search for a scaling behaviour of some quantities versus suitable variables.

In reference \cite{pion_spectrum}, the authors showed that the pion $p_{\rm T}$ spectra at midrapidity in Au-Au collisions with a center-of-mass energy per nucleon pair ($\sqrt{s_{\rm NN}}$) 200 GeV at the Relativistic Heavy Ion Collider (RHIC) presented a scaling behaviour independent of the centrality of the collisions. It was exhibited when the spectra were expressed as a function of $z=p_{\rm T}/K$, where $K$ is the scaling parameter relying on the centrality. This scaling behaviour was extended to non-central regions in Au-Au and d-Au collisions \cite{non_central_collision}. For protons and antiprotons, there was a similar scaling behaviour in their spectra in Au-Au collisions at RHIC \cite{proton_antiproton_spectra}.

In our earlier works, we observed a scaling behaviour in the $p_{\rm T}$ spectra of inclusive charged hadrons, identified charged hadrons (pions, kaons and protons) and strange hadrons ($K_{S}^{0}$, $\rm \Lambda$, $\rm \Xi$ and $\phi$) in proton-proton (pp) collisions at the Large Hadron Collider (LHC) \cite{inclusive_scaling, pi_k_p_scaling, strange_scaling}. This scaling behaviour was independent of the center of mass energy ($\sqrt{s}$). It appeared when the spectra were presented in terms of  $z=p_{\rm T}/K$, where the scaling parameter $K$ depends on $\sqrt{s}$.  We showed that the scaling behaviour in the pion, kaon, proton and the strange hadron spectra could be explained by the colour string percolation (CSP) model. The primary collisions performed at the LHC were pp collisions. However, in order to investigate the quark-gluon plasma (QGP), there were also lead-lead (Pb-Pb) collisions. Recently, the ALICE collaboration at the LHC have published data on the pion, kaon and proton $p_{\rm T}$ spectra for six different centralities  (0-5$\%$, 5-10$\%$, 10-20$\%$, 20-40$\%$, 40-60$\%$ and 60-80$\%$) in Pb-Pb collisions at $\sqrt{s_{\rm NN}}=$ 2.76 TeV \cite{Pb_Pb_collisions}. The center-of-mass energy per nucleon pair of Pb-Pb collisions at the LHC is much larger than that of Au-Au collisions at RHIC. As the scaling behaviour independent of the centrality of collisions was observed in the pion and proton $p_{\rm T}$ spectra in Au-Au collisions at $\sqrt{s_{\rm NN}}=$ 200 GeV, it is interesting to check whether a similar scaling behaviour is exhibited in the pion, kaon and proton $p_{\rm T}$ spectra in Pb-Pb collisions at $\sqrt{s_{\rm NN}}=$ 2.76 TeV. If the scaling behaviour exists, then one would like to ask two questions: (1) Can the CSP model utilized for the explanation of the scaling behaviour in pp collisions be adopted to explain the scaling behaviour of the pion, kaon and proton spectra in Pb-Pb collisions? (2) Is the dependence of the scaling parameter $K$ on centralities of the pion, kaon and proton spectra in Pb-Pb collisions the same as that in Au-Au collisions?

The organization of this paper is as follows. In section \ref{sec:method}, we will illustrate the method to search for the scaling behaviour in the pion, kaon and proton $p_{\rm T}$ spectra. In section \ref{sec:scaling_behaviour}, we will present the scaling behaviour. In section \ref{sec:mechanism}, we will discuss the scaling behaviour in the framework of the CSP model. The conclusion is drawn in section \ref{sec:conclusion}.

\section{Method to search for the scaling behaviour} \label{sec:method}

As done in reference \cite{proton_antiproton_spectra}, we will search for the scaling behaviour in the pion, kaon and proton $p_{\rm{T}}$ spectra at different centralities through the following steps. Let us take the pion $p_{\rm{T}}$ spectra as an example. First, we define a scaling variable, $z=p_{\rm{T}}/K$, and a scaled $p_{\rm{T}}$ spectrum, $\Phi(z)=A(\langle N_{\rm{part}}\rangle 2 \pi p_{\rm{T}})^{-1}d^{2}N/dp_{\rm{T}}dy|_{p_{\rm{T}}=Kz}$. Here $y$ is the rapidity of pions, $\langle N_{\rm{part}}\rangle$ is the average value of the number of participating nucleons, $(\langle N_{\rm{part}}\rangle 2 \pi p_{\rm T})^{-1}d^{2}N/dp_{\rm T}dy$ is the invariant yield of pions per participating nucleon. By choosing proper scaling parameters $K$ and $A$ that depend on the centrality, we try to coalesce the data points at different centralities in Pb-Pb collisions at $\sqrt{s_{\rm NN}}=$ 2.76 TeV into one curve.  As a convention, $K$ and $A$ at the most central collisions (0-5$\%$) are set to be 1. With this choice, $\Phi(z)$ is nothing but the $p_{\rm{T}}$ spectrum at this centrality. $K$ and $A$ at other five centralities (5-10$\%$, 10-20$\%$, 20-40$\%$, 40-60$\%$ and 60-80$\%$) will be determined by the quality factor (QF) method \cite{QF_1,QF_2}. Obviously, the scaling function $\Phi(z)$ relies on the choice of $K$ and $A$ at the most central collisions. The arbitrariness of $\Phi(z)$ can be removed if the spectra are expressed in terms of $u=z/\langle z \rangle=p_{\rm T}/\langle p_{\rm T} \rangle$. Here $\langle z \rangle=\int^{\infty}_{0}z\Phi(z)zdz\big/\int^{\infty}_{0}\Phi(z)zdz$. The corresponding normalized scaling function is $\Psi(u)=\langle z \rangle^{2}\Phi(\langle z \rangle u)\big/\int^{\infty}_{0}\Phi(z)zdz$. With $\Psi(u)$, the spectra at other five centralities can be parameterized as $f(p_{\rm T})=\langle N_{\rm{part}}\rangle /(A\langle z \rangle^{2}) \int^{\infty}_{0}\Phi(z)zdz\Psi \left(p_{\rm T}/(K\langle z \rangle \right))$, where $K$ and $A$ are the scaling parameters at these centralities. The methods to search for the scaling behaviour of the kaon and proton spectra are identical to that for the pion spectra.

\section{Scaling behaviour of the pion, kaon and proton $p_{\rm{T}}$ spectra}\label{sec:scaling_behaviour}
As described in section \ref{Intro},  the pion, kaon and proton $p_{\rm{T}}$ spectra for six different centralities covering 0-80$\%$ in Pb-Pb collisions at $\sqrt{s_{\rm NN}}=2.76$ TeV were published by the ALICE collaboration in reference \cite{Pb_Pb_collisions}. Here the pion, kaon and proton $p_{\rm{T}}$ spectra refer to the spectra of $\pi^{+}+\pi^{-}$, $K^{+}+K^{-}$ and $p+\bar{p}$ per participating nucleon. The average value of the number of participating nucleons ($\langle N_{\rm{part}}\rangle$) for each centrality is taken from reference \cite{N_part}. As the parameters $K$ and $A$ at the 0-5$\%$ centrality are set as 1, the scaling function $\Phi(z)$ is exactly the $p_{\rm{T}}$ spectrum of pions, kaons or protons at this centrality. In reference \cite{fit_function}, the authors argued the charged hadrons produced in Pb-Pb collisions stem either from QGP or from jets. The hadron yield from QGP is referred to as the soft yield, while the yield from jets is referred to as the hard yield. Both the soft and hard yields are assumed to be Tsallis distributions \cite{Tsallis_distribution}. The charged hadron spectra in Pb-Pb collisions then can be expressed using the double-Tsallis formula:
\begin{eqnarray}
E\frac{d^{3}N}{dp^{3}} = C_{1}\left(1+\frac{E_{\rm{T}}}{n_{1}T_{1}}\right)^{-n_{1}}+C_{2}\left(1+\frac{E_{\rm{T}}}{n_{2}T_{2}}\right)^{-n_{2}},
\label{eq:Tsallis_distribution_Pb-Pb}
\end{eqnarray}
where $C_{1}$, $n_{1}$, $T_{1}$, $C_{2}$, $n_{2}$ and $T_{2}$ are free parameters, $E_{\rm{T}} = m_{\rm{T}} - m = \sqrt{m^{2}+p_{\rm{T}}^{2}} - m$, $m$ is the charged hadron mass. The first Tsallis distribution represents the soft yield, while the second Tsallis distribution represents the hard yield. We find that the formula in equation (\ref{eq:Tsallis_distribution_Pb-Pb}) is able to describe the pion and kaon spectra, but fails to depict the proton spectra in Pb-Pb collisions. In order to describe the pion, kaon and proton spectra at the same time, as done in reference \cite{flow_fit_funtion}, we take the transverse flow of particles in a co-moving system into account and modify the first Tsallis distribution in equation (\ref{eq:Tsallis_distribution_Pb-Pb}) as $h(p_{\rm T}) = C_{1}\left(1+(\gamma(m_{\rm{T}}-\beta p_{\rm{T}})-m)/(n_{1}T_{1})\right)^{-n_{1}}$, where $\gamma = 1/\sqrt{1-\beta^{2}}$ is the Lorentz factor and $\beta$ is the average transverse velocity of the system.  Here we fix $\beta$ to 0.651, which is the average value of the transverse flow velocity returned by the combined blast-wave fit to the pion, kaon and proton spectra at the 0-5$\%$ centrality in the ranges 0.5-1 GeV/c, 0.2-1.5 GeV/c and 0.3-3 GeV/c respectively  \cite{pt_3_mean}. With the modified double-Tsallis distribution, the scaling function $\Phi(z)$ can be parameterized as
\begin{eqnarray}
\Phi(z)=C_{1}\left(1+\frac{\gamma(\sqrt{m^{2}+z^{2}}-\beta z)-m}{n_{1}T_{1}}\right)^{-n_{1}}+C_{2}\left(1+\frac{\sqrt{m^{2}+z^{2}}-m}{n_{2}T_{2}}\right)^{-n_{2}}.
\label{eq:phi_z_pt_spectrum_PbPb}
\end{eqnarray}
The free parameters $C_{1}$, $n_{1}$, $T_{1}$, $C_{2}$, $n_{2}$ and $T_{2}$ are determined by fitting equation (\ref{eq:phi_z_pt_spectrum_PbPb}) to the pion, kaon and proton $p_{\rm{T}}$ spectra at the 0-5$\%$ centrality separately with the least $\chi^{2}$s method. The statistical and systematic uncertainties of the data points have been added in quadrature in the fits. Table \ref{tab:id_particles_fit_parameters} lists the parameters $C_{1}$, $n_{1}$, $T_{1}$, $C_{2}$, $n_{2}$, $T_{2}$ and their uncertainties returned by the fits. The last line of this table shows the $\chi^2$s per degrees of freedom (dof), named reduced $\chi^{2}$s, for these fits. As the systematic uncertainties in the data points are correlated and relatively large, the reduced $\chi^{2}$s for the fits on the pion, kaon and proton spectra are relatively low. For protons, the value of $n_{1}$  (9.7$\times 10^{4}$) returned from the fit is very large. As described in reference \cite{Tsallis_distribution_1}, when $n_{1}$ tends to be infinity, $h(p_{\rm T})$ tends to be an exponential distribution, $h(p_{\rm T}) = C_{1}\exp(-(\gamma(m_{\rm{T}}-\beta p_{\rm{T}})-m)/T_{1})$. Thus we set $n_{1}$ to be infinity and redo the fit to the proton spectrum at the 0-5$\%$ centrality. The fit parameters are listed in the fourth column of table \ref{tab:id_particles_fit_parameters}.

\begin{table}[H]
  \caption{\label{tab:id_particles_fit_parameters} $C_{1}$, $n_{1}$, $T_{1}$, $C_{2}$, $n_{2}$ and $T_{2}$ of $\Phi(z)$ for pions, kaons and protons. The uncertainties quoted are due to the statistical plus systematic errors of the data points added in quadrature. The last line shows the reduced $\chi^{2}$s for the fits on the pion, kaon and proton $p_{\rm T}$ spectra at the 0-5$\%$ centrality in Pb-Pb collisions at $\sqrt{s_{\rm NN}}=2.76$ TeV.}
\begin{center}
\begin{tabular}{@{}cccc}
\toprule
 \textrm{\ }&
\textrm{Pions}&
\textrm{Kaons}&
\textrm{Protons}
\\
\hline
$C_{1}$ &1.93$\pm$0.03&0.106$\pm$0.002&0.0144$\pm$0.0002\\
$n_{1}$ & 21.2$\pm$0.4 & 27.20$\pm$2.15&          $\infty$\\
$T_{1}$ & 0.1206$\pm$0.0006 &0.142$\pm$0.002 &0.182$\pm$0.001\\
$C_{2}$ & 9.1$\pm$0.1 & 0.200$\pm$0.009 & 0.015$\pm$0.001\\
$n_{2}$ & 5.83$\pm$0.02 & 5.64$\pm$0.07 &  7.7$\pm$0.3\\
$T_{2}$ & 0.0889$\pm$0.0007 & 0.142$\pm$0.003 & 0.26$\pm$0.01\\
\hline
$\chi^{2}$/dof & 1.43/57 & 6.84/52 & 13.33/44\\
\toprule
\end{tabular}
\end{center}
\end{table}

As described in section 2, $K$ and $A$ at 5-10$\%$, 10-20$\%$, 20-40$\%$, 40-60$\%$ and 60-80$\%$ centralities will be determined by the QF method. With this method, the dependence of $K$ and $A$ on the shape of the scaling function will be eliminated. Thus, compared with the method utilized in references \cite{proton_antiproton_spectra, inclusive_scaling}, the QF method is more robust. Given a set of data points ($\rho^{i}, \tau^{i}$), where $\rho^{i}=p_{\rm T}^{i}/K$ and $\tau^{i}=\textrm{log}(A(\langle N_{\rm{part}}\rangle 2\pi p^{i}_{\rm T})^{-1}d^{2}N^{i}/dp^{i}_{\rm T}dy^{i})$, the QF is defined as follows \cite{QF_1,QF_2}
\begin{eqnarray}
\textrm{QF}(K,A)=\left[\sum_{i=2}^{n}\frac{(\tau^{i}-\tau^{i-1})^{2}}{(\rho^{i}-\rho^{i-1})^{2}+1/n^{2}}\right]^{-1},
\label{eq:QF_definition}
\end{eqnarray}
where $n$ is the number of data points and $1/{n^{2}}$ is utilized to keep the sum finite when two points have the same $\rho$ value.  Before entering the QF, $\rho^{i}$ are ordered, $\tau^{i}$ are rescaled so that they are in the range between 0 and 1. It is obvious that a large contribution to the sum in the QF is given if two successive data points are close in $\rho$ and far in $\tau$. Therefore, a set of data points are expected to lie close to a single curve if they have a small sum (a large QF) in equation (\ref{eq:QF_definition}). The best set of ($K$, $A$) at the 5-10$\%$ (10-20$\%$, 20-40$\%$, 40-60$\%$ and 60-80$\%$) centrality is chosen to be the one which globally maximizes the QF of data points at 5-10$\%$ (10-20$\%$, 20-40$\%$, 40-60$\%$ and 60-80$\%$) and 0-5$\%$ centralities. Table \ref{tab:id_particles_a_k_parameters} tabulates $K$ and $A$ for the pion, kaon and proton $p_{\rm T}$ spectra at 5-10$\%$, 10-20$\%$, 20-40$\%$, 40-60$\%$ and 60-80$\%$ centralities. Also presented in the table are the maximum quality factors (QF$_{\rm max}$). Here we have set the $K$ values at these centralities to be less than 1. The reason is as follows. If the scaling behaviour in the pion, kaon and proton $p_{\rm T}$ spectra exist, then the values of $\langle z \rangle$ at different centralities are the same. Since $K= \langle p_{\rm T}\rangle/\langle z \rangle$, $K$ should be proportional to the mean $p_{\rm T}$. As shown in reference \cite{pt_3_mean}, the more central the collisions are, the larger the $\langle p_{\rm{T}} \rangle$ is. Thus the $K$ values at 5-10$\%$, 10-20$\%$, 20-40$\%$, 40-60$\%$ and 60-80$\%$ centralities are smaller than the value at the 0-5$\%$ centrality.

\begin{figure}[H]
\centering
\includegraphics[scale=0.17]{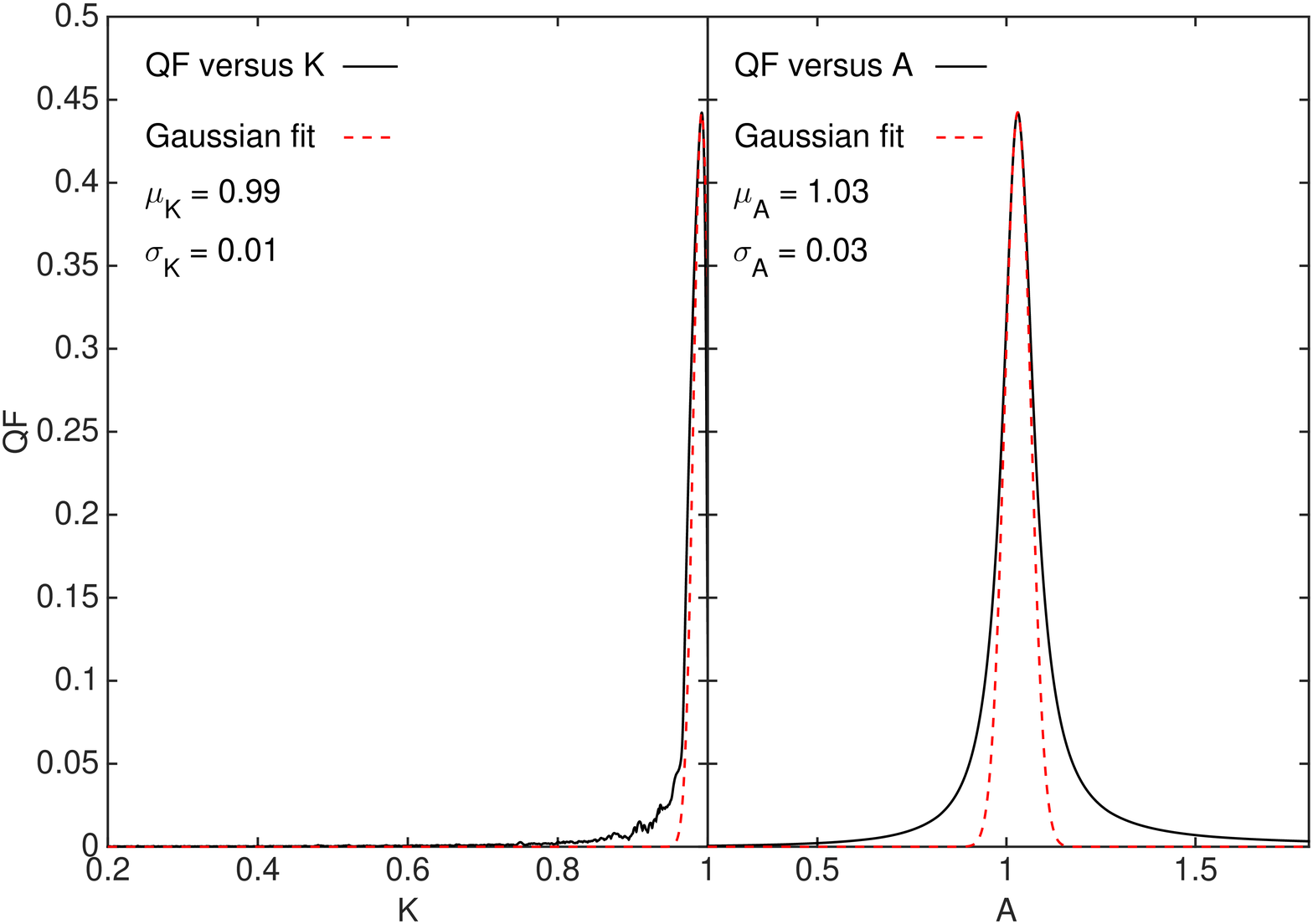}
\caption{\label{fig:QF_vs_a_k_pion_0_5_cent}The left (right) panel: QF versus $K$ ($A$) for the pion spectrum at the 0-5$\%$ centrality, with $A$ ($K$) fixed to 1.03 (0.99). The black solid curve is the QF scatter plot, the red dash curve is the Gaussian fit of the peak with $\rm QF>0.43$.}
\end{figure}

In order to evaluate the uncertainties of $K$ and $A$, we adopt the method in reference \cite{QF_1}. Taking the determination of the $K$ and $A$ errors for the pion spectrum at the 5-10$\%$ centrality as an example, in figure \ref{fig:QF_vs_a_k_pion_0_5_cent} we plot the QF as a function of $K$ ($A$) with $A$ ($K$) fixed to the value 1.03 (0.99) returned by the QF method. The peak value with $\rm QF >(QF_{max}-0.01)$ shows a good scaling and we make a Gaussian fit to this bump. The standard deviation of the Gaussian fit, $\sigma_{K(A)}$, is taken as the uncertainty of $K$ ($A$) for the pion spectrum at the 5-10$\%$ centrality. The mean value of the Gaussian fit, $\mu_{K(A)}$, is consistent with the value of $K$ ($A$) returned by the QF method, thus the method to determine the errors of $K$ and $A$ is robust. The $K$ and $A$ errors for the pion, kaon and proton spectra at 10-20$\%$, 20-40$\%$, 40-60$\%$ and 60-80$\%$ centralities are determined by making Gaussian fits to their QF peaks with $\rm QF>(QF_{max}-0.01)$.

\begin{table}[H]
  \caption{\label{tab:id_particles_a_k_parameters} $K$ and $A$ for the pion, kaon and proton spectra at 5-10$\%$, 10-20$\%$, 20-40$\%$, 40-60$\%$ and 60-80$\%$ centralities. The $\rm QF_{max}$ is shown in the last column of the table. The standard deviations of the Gaussian fits to the peaks of the QF scatter plots are taken as the uncertainties of $K$ and $A$ at these centralities.}
\begin{center}
\begin{tabular}{@{}cccccc}
\toprule
 \textrm{\ }&
\textrm{Centrality}&
\textrm{$K$}&
\textrm{$A$}&
$\rm QF_{max}$\\
\hline
\textrm{\ }&5-10$\%$&0.99$\pm$0.01&1.03$\pm$0.03&0.44\\
\textrm{\ }&10-20$\%$&0.99$\pm$0.01&1.08$\pm$0.05&0.35\\
\textrm{Pions}&20-40$\%$&0.98$\pm$0.01&1.16$\pm$0.07&0.26\\
\textrm{\ }&40-60$\%$&0.98$\pm$0.01&1.26$\pm$0.10&0.16\\
\textrm{\ }&60-80$\%$&0.96$\pm$0.03&1.39$\pm$0.14&0.09\\
\hline
\textrm{\ }&5-10$\%$&0.99$\pm$0.01&1.02$\pm$0.04&0.85\\
\textrm{\ }&10-20$\%$&0.98$\pm$0.01&1.06$\pm$0.05&0.61\\
\textrm{Kaons}&20-40$\%$&0.98$\pm$0.01&1.11$\pm$0.08&0.40\\
\textrm{\ }&40-60$\%$&0.96$\pm$0.02&1.21$\pm$0.10&0.23\\
\textrm{\ }&60-80$\%$&0.96$\pm$0.02&1.62$\pm$0.21&0.09\\
\hline
\textrm{\ }&5-10$\%$&0.99$\pm$0.01&1.03$\pm$0.03&2.20\\
\textrm{\ }&10-20$\%$&0.98$\pm$0.01&1.04$\pm$0.04&1.71\\
\textrm{Protons}&20-40$\%$&0.94$\pm$0.01&0.99$\pm$0.05&0.74\\
\textrm{\ }&40-60$\%$&0.86$\pm$0.02& 0.88$\pm$0.08&0.24\\
\textrm{\ }&60-80$\%$&0.751$\pm$0.004& 0.75$\pm$0.11&0.12\\
\toprule
\end{tabular}
\end{center}
\end{table}

\begin{figure}[H]
\centering
\includegraphics[scale=0.17]{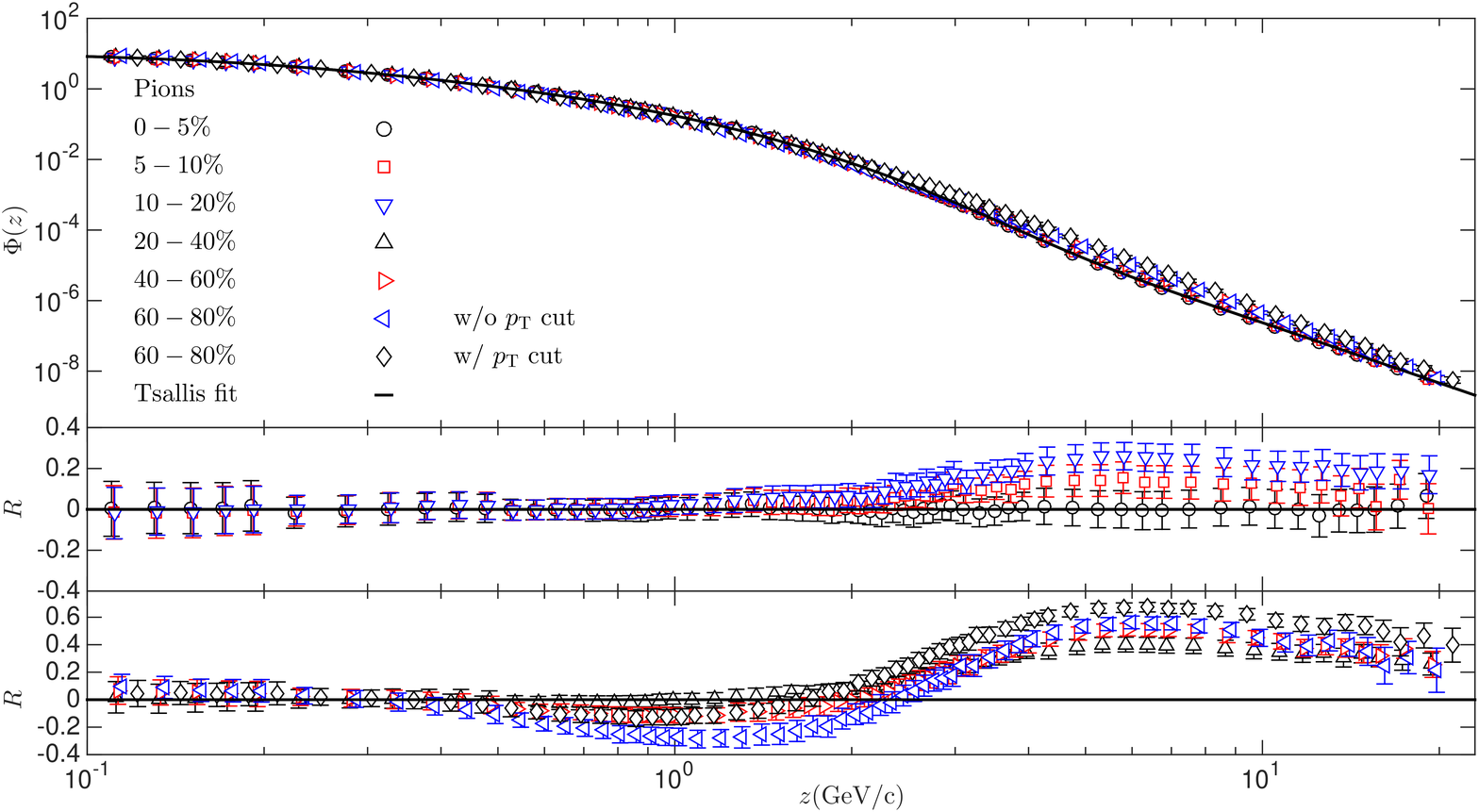}
\caption{\label{fig:pi_z_plus_ratio}The upper panel: the scaling behaviour in the pion $p_{\rm{T}}$ spectra presented in $z$ at 0-5$\%$, 5-10$\%$, 10-20$\%$, 20-40$\%$, 40-60$\%$ and 60-80$\%$ centralities. The black curve is described by equation (\ref{eq:phi_z_pt_spectrum_PbPb}) with parameters in the second column of table \ref{tab:id_particles_fit_parameters}.  The data points are taken from reference \cite{Pb_Pb_collisions}. The middle (lower) panel: the $R$ distributions at 0-5$\%$, 5-10$\%$ and 10-20$\%$ (20-40$\%$, 40-60$\%$ and 60-80$\%$) centralities. }
\end{figure}

With $K$ and $A$ in table \ref{tab:id_particles_a_k_parameters}, now we try to put the pion $p_{\rm T}$ spectra at 5-10$\%$, 10-20$\%$, 20-40$\%$, 40-60$\%$ and 60-80$\%$ centralities to the spectrum at the 0-5$\%$ centrality. As shown in the upper panel of figure \ref{fig:pi_z_plus_ratio}, on a log scale most of the data points at different centralities appear consistent with the universal curve described by the scaling function $\Phi(z)$ in equation (\ref{eq:phi_z_pt_spectrum_PbPb}) with parameters in the second column of table \ref{tab:id_particles_fit_parameters}. In order to see the agreement between the experimental data and the fitted curve, we define a ratio $R=\rm (data-fitted)/data$ and plot the $R$ distributions for the spectra at 0-5$\%$, 5-10$\%$ and 10-20$\%$ (20-40$\%$, 40-60$\%$ and 60-80$\%$) centralities in the middle (lower) panel of this figure. The error bars represent the uncertainties of $R$, $\Delta R = \rm (fitted/data)(\Delta data/data)$, where $\rm \Delta data$ is the total uncertainty of the data point. For the spectra at 0-5$\%$ and 5-10$\%$ centralities, all the data points have $R$ values from -0.2 to 0.2. For the spectra at the 10-20$\%$, 20-40$\%$ and 40-60$\%$ centralities, the data points in the regions with $z \leq$ 3.74, 2.80 and 2.92 ($p_{\rm T} \leq$ 3.70, 2.75 and 2.85) GeV/c have absolute $R$ values less than 0.2. Thus, in the region with $z \leq$ 2.80 ($p_{\rm T} \leq$ 2.75) GeV/c, all the data points at the first five centrality bins agree with the fitted curve within 20$\%$. However, in this region, for the spectrum at the 60-80$\%$ centrality there are some data points with $0.70\leq z \leq 1.61$ GeV/c differing significantly from the fitted curve. In order to improve this agreement, we apply the cut $p_{\rm T} \leq$ 2.75 GeV/c to the spectra at the 0-5$\%$ and 60-80$\%$ centralities and utilize the QF method to determine a new set of ($K$, $A$). It is (0.90$\pm$0.01, 1.27$\pm$0.13). With this new set of ($K$, $A$), we try to rescale the whole $p_{\rm T}$ spectrum at the 60-80$\%$ centrality in figure \ref{fig:pi_z_plus_ratio}. The $R$ distribution (marked as the diamond symbol) is shown in the lower panel of this figure. From this distribution we see that in the range with $z \leq 2.80$ GeV/c, except for the last four points, all the other data points at the 60-80$\%$ centrality agree with the fitted curve within 20$\%$.

\begin{figure}[H]
\centering
\includegraphics[scale=0.17]{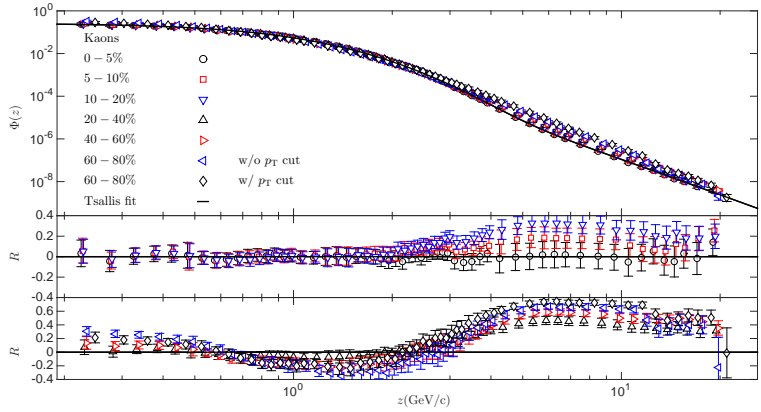}
\caption{\label{fig:k_z_plus_ratio} The upper panel: the scaling behaviour in the kaon $p_{\rm{T}}$ spectra presented in $z$ at 0-5$\%$, 5-10$\%$, 10-20$\%$, 20-40$\%$, 40-60$\%$ and 60-80$\%$ centralities. The black curve is described by equation (\ref{eq:phi_z_pt_spectrum_PbPb}) with parameters in the third column of table \ref{tab:id_particles_fit_parameters}.  The data points are taken from reference \cite{Pb_Pb_collisions}. The middle (lower) panel: the $R$ distributions at 0-5$\%$, 5-10$\%$ and 10-20$\%$ (20-40$\%$, 40-60$\%$ and 60-80$\%$) centralities.}
\end{figure}

In figure \ref{fig:k_z_plus_ratio}, we present the scaling behaviour of the kaon spectra at different centralities. For the spectra at 0-5$\%$, 5-10$\%$ and 10-20$\%$ centralities, except for the last point at the 5-10$\%$ centrality and few points with $3.76 \leq z \leq 14.73$  GeV/c at the 10-20$\%$ centrality, all the other data points are consistent with the fitted curve within 20$\%$.  For the spectrum at the 20-40$\%$ centrality, all the data points in the range with $z \leq$ 3.17 ($p_{\rm T} \leq$ 3.10) GeV/c agree with the fitted curve within 20$\%$. For the spectrum at the 40-60$\%$ centrality, except for three points with $z=$1.19, 1.30 and 1.40 GeV/c, all the other data points in the range with $z \leq$ 3.22 ($p_{\rm T} \leq$ 3.10) GeV/c have absolute $R$ values less than 0.2. Thus, in the region with $z \leq$ 3.17 ($p_{\rm T} \leq$ 3.10) GeV/c, almost all the data points at the first five centralities agree with the fitted curve within 20$\%$. However, in this region, for the spectrum at the 60-80$\%$ centrality, there are a few points with $z \leq 0.39$ GeV/c and $0.96\leq z \leq 2.03$ GeV/c having absolute $R$ values greater than 0.2. In order to improve the agreement between the data points at this centrality and the fitted curve, we apply the cut $p_{\rm T} \leq$ 3.10 GeV/c to the spectra at the 0-5$\%$ and 60-80$\%$ centralities and utilize the QF method to determine a new set of ($K$, $A$). It is (0.91$\pm$0.01, 1.42$\pm$0.16). With this new set of ($K$, $A$), we try to re-shift the whole $p_{\rm T}$ spectrum at the 60-80$\%$ centrality to the spectrum at the 0-5$\%$ in figure \ref{fig:k_z_plus_ratio}. The $R$ distribution (marked as the diamond symbol) is shown in the lower panel of this figure. From this distribution we see that in the range with $z \leq 3.17$ GeV/c, except for the last four points and the points with $z=$ 0.25, 1.27, 1.38 and 1.49 GeV/c, all the other data points at the 60-80$\%$ centrality agree with the fitted curve within 20$\%$.

\begin{figure}[H]
\centering
\includegraphics[scale=0.17]{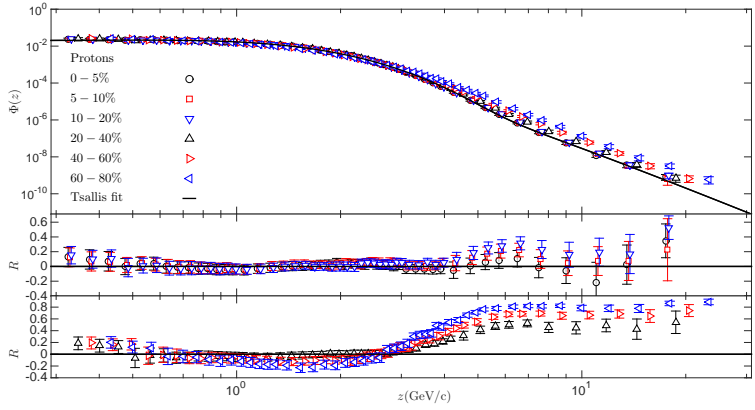}
\caption{\label{fig:p_z_plus_ratio}Upper panel: the scaling behaviour in the proton $p_{\rm{T}}$ spectra presented in $z$ at 0-5$\%$, 5-10$\%$, 10-20$\%$, 20-40$\%$, 40-60$\%$ and 60-80$\%$ centralities. The black curve is described by equation (\ref{eq:phi_z_pt_spectrum_PbPb}) with parameters listed in the fourth column of table \ref{tab:id_particles_fit_parameters}.  The data points are taken from reference \cite{Pb_Pb_collisions}. The middle (lower) panel: the $R$ distributions at 0-5$\%$, 5-10$\%$ and 10-20$\%$ (20-40$\%$, 40-60$\%$ and 60-80$\%$) centralities. }
\end{figure}

The scaling behaviour of the proton spectra is shown in figure \ref{fig:p_z_plus_ratio}. For the spectrum at 0-5$\%$ centrality, except for the last point, all the other data points have absolute values of $R$ less than 0.2. For the spectra at 5-10$\%$, 10-20$\%$, 20-40$\%$ and 40-60$\%$ centralities, the data points in the regions with $z \leq$ 5.31, 4.85, 3.74 and 3.21($p_{\rm T} \leq$ 5.25, 4.75, 3.50 and 2.75) GeV/c are consistent with the fitted curve within 20$\%$. For the spectrum at the 60-80$\%$ centrality, except for two points at $z=$ 1.53 and 1.66 GeV/c, all the other data points have absolute $R$ values smaller than 0.2 in the range with $z \leq 3.13$ ($p_{\rm T} \leq$ 2.35) GeV/c. Thus, in the region with $z \leq 3.13$ ($p_{\rm T} \leq$ 2.35) GeV/c, almost all the data points at each centrality agree with the fitted curve within 20$\%$.

So far, we have seen that the pion, kaon and proton spectra at all centralities exhibit a scaling behaviour in the low $p_{\rm T}$ regions with $z \leq$ 2.80, 3.17 and  3.13 ($p_{\rm T} \leq$ 2.75, 3.10 and 2.35) GeV/c. In these regions, almost all the data points agree with the fitted curves within 20$\%$. Outside these regions, as the deviations between the data points and the fitted curve are larger than the experimental uncertainties, a violation of the proposed scaling is observed going form central to peripheral collisions. The more peripheral the collisions are, the more clearly violated the proposed scaling becomes. This is due to reason that there is a suppression of the spectra in the large $p_{\rm T}$ region. The more central the collisions are, the larger the suppression is \cite{spectra_suppression}.

\begin{figure}[h]
\centering
\includegraphics[scale=0.17]{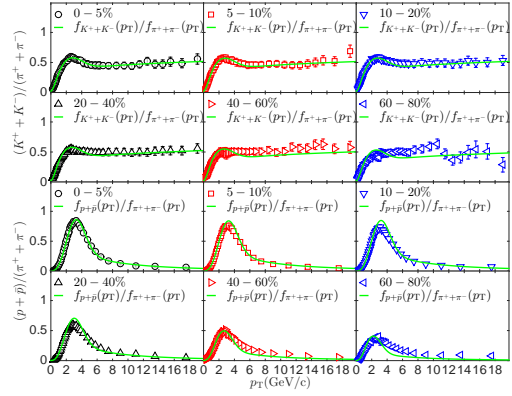}
\caption{\label{fig:kaon_div_pion_and_p_div_pion} The panels in the top (bottom) two rows: $(K^{+}+K^{-})/(\pi^{+}+\pi^{-})$ ($(p+\bar{p})/(\pi^{+}+\pi^{-})$) distributions at 0-5$\%$, 5-10$\%$, 10-20$\%$, 20-40$\%$, 40-60$\%$ and 60-80$\%$ centralities. The green curves are $f_{K^{+}+K^{-}}(p_{\rm{T}})/f_{\pi^{+}+\pi^{-}}(p_{\rm{T}})$ and $f_{p+\bar{p}}(p_{\rm{T}})/f_{\pi^{+}+\pi^{-}}(p_{\rm{T}})$  at these centralities. The data points are taken from reference \cite{Pb_Pb_collisions}.}
\end{figure}

As described in the first paragraph of this section, the scaling function $\Phi(z)$ could be written as the modified double-Tsallis distribution with the transverse flow velocity taken into accounted in equation (\ref{eq:phi_z_pt_spectrum_PbPb}) phenomenologically. As shown in reference \cite{pt_3_mean}, the transverse velocity extracted from the combined blast-wave fit to the pion, kaon and proton spectra depends on the centrality. In this work, the conclusion drawn from the scaling behaviour of the pion, kaon and proton spectra is that the transverse velocities of the scaled (not the original) spectra in the low $p_{\rm T}$ region at the 5-10$\%$, 10-20$\%$, 20-40$\%$, 40-60$\%$ and 60-80$\%$ centralities are the same as the one at the 0-5$\%$ centrality. Thus this conclusion is not in contradiction with the result from reference \cite{pt_3_mean}. Since the scaling function $\Phi(z)$ relies on the choice of $K$ and $A$ at the 0-5$\%$ centrality. In order to eliminate this reliance, we replace $z$ with another scaling variable $u = z/\langle z \rangle$. The $\langle z \rangle$ values for the pion, kaon and proton $p_{\rm{T}}$ spectra are $0.52\pm0.02$, $0.88\pm0.03$ and $1.35\pm0.03$ GeV/c. With the substitution of $\langle z \rangle$ and $\Phi(z)$ into $\Psi(u)$, the normalized scaling function can be written as
\begin{eqnarray}
\Psi(u)=C^{\prime}_{1}\left(1+\frac{\gamma(\sqrt{m^{\prime 2}+u^{2}}-\beta u)-m^{\prime}}{n_{1}T^{\prime}_{1}}\right)^{-n_{1}}+C^{\prime}_{2}\left(1+\frac{\sqrt{m^{\prime 2}+u^{2}}-m^{\prime}}{n_{2}T^{\prime}_{2}}\right)^{-n_{2}},
\label{eq:psi_u_pt_spectrum_PbPb}
\end{eqnarray}
where $C^{\prime}_{1,2}=\langle z \rangle^{2}C_{1,2}/\int^{\infty}_{0}\Phi(z)zdz$, $T^{\prime}_{1,2}=T_{1,2}/\langle z \rangle$ and $m^{'}=m/\langle z \rangle$. Table \ref{tab:id_particles_normalized_parameters} lists their values. With $\Psi(u)$, the pion, kaon and proton spectra at 5-10$\%$, 10-20$\%$, 20-40$\%$, 40-60$\%$ and 60-80$\%$ centralities are parameterized as $f(p_{\rm T})=\langle N_{\rm{part}}\rangle /(A\langle z \rangle^{2}) \int^{\infty}_{0}\Phi(z)zdz\Psi \left(p_{\rm T}/(K\langle z \rangle \right))$, where $K$ and $A$ are the scaling parameters at these centralities. In reference \cite{Pb_Pb_collisions}, the ALICE collaboration have presented the kaon to pion ratio ($(K^{+}+K^{-})/(\pi^{+}+\pi^{-})$) and the proton to pion ratio ($(p+\bar{p})/(\pi^{+}+\pi^{-})$) as a function of $p_{\rm{T}}$ at different centralities. Here in figure \ref{fig:kaon_div_pion_and_p_div_pion} we show in the low $p_{\rm{T}}$ region the $(K^{+}+K^{-})/(\pi^{+}+\pi^{-})$ ($(p+\bar{p})/(\pi^{+}+\pi^{-})$) distributions at 5-10$\%$, 10-20$\%$, 20-40$\%$, 40-60$\%$ and 60-80$\%$ centralities are well described by $f_{K^{+}+K^{-}}(p_{\rm{T}})/f_{\pi^{+}+\pi^{-}}(p_{\rm{T}})$ ($f_{p+\bar{p}}(p_{\rm{T}})/f_{\pi^{+}+\pi^{-}}(p_{\rm{T}})$). This agreement is a definite indication that the scaling behaviour exists in the low $p_{\rm{T}}$ region of the spectra. In the high $p_{\rm{T}}$ region, due to the violation of the scaling, there is an obvious deviation of $f_{K^{+}+K^{-}}(p_{\rm{T}})/f_{\pi^{+}+\pi^{-}}(p_{\rm{T}})$ ($f_{p+\bar{p}}(p_{\rm{T}})/f_{\pi^{+}+\pi^{-}}(p_{\rm{T}})$) from the data points at peripheral collisions.

\begin{table}[H]
\caption{\label{tab:id_particles_normalized_parameters} $C^{\prime}_{1}$, $T^{\prime}_{1}$, $C^{\prime}_{2}$, $T^{\prime}_{2}$  and $m^{'}$ of the normalized scaling functions $\Psi(u)$ for the pion, kaon and proton spectra. The uncertainties quoted are due to the error of $C_{1}$, $n_{1}$, $T_{1}$, $C_{2}$, $n_{2}$ and $T_{2}$ in table \ref{tab:id_particles_fit_parameters}.}
\begin{center}
\begin{tabular}{@{}cccc}
\toprule
 \textrm{\ }&
\textrm{Pions}&
\textrm{Kaons}&
\textrm{Protons}
\\
\hline
$C^{'}_{1}$ & 0.83$\pm$0.07 & 0.90$\pm$0.06 &0.95$\pm$0.04\\
$T^{'}_{1}$ & 0.234$\pm$0.009 & 0.161$\pm$0.006 & 0.134$\pm$0.003\\
$C^{'}_{2}$ & 3.934$\pm$0.3 & 1.702$\pm$0.1 & 0.960$\pm$0.08\\
$T^{'}_{2}$ & 0.172$\pm$0.007& 0.162$\pm$0.007& 0.19$\pm$0.01\\
$m^{'}$     & 0.271$\pm$0.005 & 0.56$\pm$0.02 & 0.69$\pm$0.02\\
\toprule
\end{tabular}
\end{center}
\end{table}

\section{Discussions}\label{sec:mechanism}
In section \ref{sec:scaling_behaviour}, we have presented that in the low $p_{\rm T}$ regions with  $z \leq$ 2.80, 3.17 and  3.13 ($p_{\rm T} \leq$ 2.75, 3.10 and 2.35) GeV/c the pion, kaon and proton spectra exhibit a scaling behaviour which is independent of the centrality. In this section, we would like to discuss this scaling behaviour in the framework of CSP model \cite{string_perco_model_1, string_perco_model_2}.

In this model, colour strings are stretched between the partons of the projectile and target in Pb-Pb collisions. These strings will decay into new ones with the emission of $q\bar{q}$ pairs. Charged hadrons such as pions, kaons and protons are produced by the hadronization of these new strings. In the transverse plane, the colour strings look like discs with areas of $S_{1}=\pi r_{0}^{2}$, where $r_{0}\approx 0.2$ fm. As described in reference \cite{string_perco_model_1}, when the centrality increases, the number of strings grows. They start to overlap with each other, which leads to the formation of clusters. The average $p_{\rm T}^{2}$, $\langle p_{\rm{T}}^{2}\rangle_{n}$, of charged pions, kaons and protons produced by a cluster with $n$ ($n>1$) strings is given by $\langle p_{\rm{T}}^{2}\rangle_{n}=\sqrt{nS_{1}/S_{n}} \langle p_{\rm{T}}^{2}\rangle_{1}$, where $\langle p_{\rm{T}}^{2}\rangle_{1}$ is the mean $p_{\rm{T}}^{2}$ of charged hadrons produced by a single string, $S_{n}$ is the transverse area of the cluster and $nS_{1}/S_{n}$ is the degree of string overlap. For the case where strings just get in touch with each other, $S_{n}=nS_{1}$, $nS_{1}/S_{n}=1$ and $\langle p_{\rm{T}}^{2}\rangle_{n}=\langle p_{\rm{T}}^{2}\rangle_{1}$, which means that $n$ strings fragment into charged hadrons independently. For the case where strings maximally overlap with each other, $S_{n}=S_{1}$, $nS_{1}/S_{n}=n$ and $\langle p_{\rm{T}}^{2}\rangle_{n}=\sqrt{n} \langle p_{\rm{T}}^{2}\rangle_{1}$, which means that due to the percolation the $\langle p_{\rm{T}}^{2}\rangle$ is maximally enhanced.

The $p_{\rm{T}}$ spectra of charged hadrons produced in Pb-Pb collisions can be expressed as a superposition of the $p_{\rm{T}}$ spectra produced by each cluster, $g(x, p_{\rm{T}})$, weighted with the cluster's size distribution, $W(x)$,
\begin{eqnarray}
\frac{d^{2}N}{2\pi p_{\rm{T}}dp_{\rm{T}}dy}=C\int_{0}^{1/p_{\rm T}^{2}}W(x)g(x,p_{\rm T})dx,
\label{eq:CPS_formula}
\end{eqnarray}
where $C$ is a normalization constant referring to the total number of clusters formed for charged hadrons before hadronization. For the cluster's size distribution $W(x)$, it is chosen to be the gamma distribution,
\begin{eqnarray}
W(x)=\frac{\zeta}{\Gamma(\kappa)}(\zeta x)^{\kappa-1}\textrm{exp}(-\zeta x),
\label{eq:gamma_function}
\end{eqnarray}
where $x=1/\langle p_{\rm T}^{2}\rangle_{n}$, $\kappa$ and $\zeta$ are free parameters. $\kappa$ is related to the dispersion of the size distribution, $1/\kappa=(\langle x^{2}\rangle-\langle x\rangle^{2})/\langle x\rangle^{2}$. $\zeta$ is related to the mean $x$, $\langle x\rangle=\kappa/\zeta$. For the cluster's fragmentation function $g(x, p_{\rm T})$, as done in reference \cite{frag_function_1}, we assume it is analogous to the usual fragmentation functions from hard partons to hadrons. Thus it is written as \cite{frag_function_2}
\begin{eqnarray}
g(x, p_{\rm T})=D\xi^{\alpha}(1-\xi)^{\eta}(1+\xi)^{\theta},
\label{eq:frag_function}
\end{eqnarray}
where $\xi=p_{\rm T}\sqrt{x}$ is the fraction of the transverse momentum of the produced hadron relative to that of a cluster whose size is $x$, $D$, $\alpha$, $\eta$ and $\theta$ are free parameters. In the real fitting, the parameter $C$ in equation (\ref{eq:CPS_formula}) and $D$ in equation (\ref{eq:frag_function}) always appear as a product, thus $D$ can be absorbed into $C$. As the fraction of transverse momentum carried by the charged hadron can not be larger than 1, the upper limit of the integration in equation (\ref{eq:CPS_formula}) is set to be $1/p_{\rm T}^{2}$, rather than infinity. For different charged hadrons, $g(x, p_{\rm T})$ should be different, since they are produced from different fragmentation channels of the clusters. However, $g(x, p_{\rm T})$ should be the same for the same charged hadrons produced from clusters with different sizes $x$. In our work, it means that $g(x, p_{\rm T})$ is universal for all the centralities of the collisions.

\begin{figure}[H]
\centering
\includegraphics[scale=0.17]{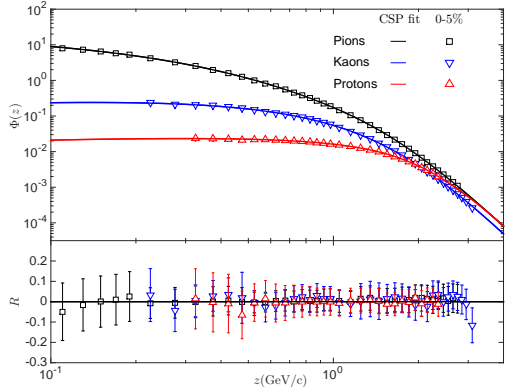}
\caption{\label{fig:all_particle_csp_fit} Upper panel: the CSP fit to the charged pion, kaon and proton spectra at the 0-5$\%$ centrality in the low $p_{\rm T}$ regions  $p_{\rm T} \leq$ 2.75, 3.10 and 2.35 GeV/c. The data points are taken from reference \cite{Pb_Pb_collisions}. Lower panel: the $R$ distributions.}
\end{figure}

As described in section \ref{sec:scaling_behaviour}, the scaling functions $\Phi(z)$ are exactly the pion, kaon and proton $p_{\rm T}$ spectra at the 0-5$\%$ centrality. In order to see whether the CSP model can describe these scaling functions in the region where the scaling behaviour is exhibited, we attempt to fit equation (\ref{eq:CPS_formula}) to the pion, kaon and proton spectra at the 0-5$\%$ centrality in the low $p_{\rm T}$ regions with $p_{\rm T} \leq$ 2.75, 3.10 and 2.35 GeV/c.  For the kaon and proton spectra, in order to make the fits on them stable, we set the parameter $\theta$ in  $g(x, p_{\rm T})$ to be 0. The parameters $C$, $\zeta$, $\kappa$,  $\alpha$, $\eta$ and $\theta$ are tabulated in table \ref{tab:CSP_fit_parameters}. From the table, we see that for different charged hadrons the cluster's fragmentation functions $g(x, p_{\rm T})$ are indeed different.  As the systematic uncertainties of the data points are correlated and relatively large, the reduced $\chi^{2}$s of the fits are rather low. The fit results are presented in the upper panel of figure \ref{fig:all_particle_csp_fit}. Almost all the data points are consistent with the CSP fits within uncertainties, which can be seen from the $R$ distributions in the lower panel of this figure.  Therefore, the scaling functions of the charged hadrons in the low $p_{\rm T}$ regions can be described by the CSP model.

Now we would like to check whether the CSP model can describe the scaling property of the charged hadron $p_{\rm T}$ spectra. Under the transformation $x \rightarrow x' = \lambda x$, $\zeta \rightarrow \zeta' = \zeta/\lambda$ and $p_{\rm{T}} \rightarrow p_{\rm{T}}' = p_{\rm{T}}/\sqrt{\lambda}$, it is obvious that both $W(x)$ and $g(x, p_{\rm T})$ are invariant. Here $\lambda = \langle S_{n}/nS_{1} \rangle^{1/2}$, where the average is taken over all the clusters decaying into charged hadrons \cite{string_perco_model_2}. As a result, the $p_{\rm{T}}$ distribution in equation (\ref{eq:CPS_formula}) is also invariant within a constant. This invariance is exactly the scaling behaviour we are looking for. Comparing the $p_{\rm{T}}'$ transformation in the CSP model $p_{\rm{T}}' \rightarrow p_{\rm{T}}'\sqrt{\lambda}$ with the one used to search for the scaling behaviour $p_{\rm{T}}\rightarrow p_{\rm{T}}/K$, we deduce that $K$ is proportional to $1/\sqrt{\lambda}$, which is equal to $\langle nS_{1}/S_{n} \rangle^{1/4}$. Since the degree of string overlap $nS_{1}/S_{n}$ grows with centralities nonlinearly \cite{string_perco_model_1}, the scaling parameter $K$ should also increase with centralities in a nonlinear trend. That's indeed what we observed for the charged hadrons in table \ref{tab:id_particles_a_k_parameters}. Therefore the CSP model can qualitatively explain the scaling behaviour for the pion, kaon and proton spectra separately.

\begin{table}[H]
  \caption{\label{tab:CSP_fit_parameters} $C$, $\zeta$, $\kappa$, $\alpha$, $\eta$ and $\theta$ of the CSP fits on the pion, kaon and proton spectra at the 0-5$\%$ centrality. The uncertainties quoted are due to the square root of the sum of the statistical and systematic uncertainties of the data points. The last line shows the reduced $\chi^{2}$s for the fits.}
\begin{center}
\begin{tabular}{@{}cccc}
\toprule
\textrm{\ }&
\textrm{Pions}&
\textrm{Kaons}&
\textrm{Protons}
\\
\hline
$C$ & 10.66$\pm$1.34 &0.73$\pm$0.24& 0.045$\pm$0.006\\
$\zeta$&8.26$\pm$0.60&21.86$\pm$5.78&21.99$\pm$3.95\\
$\kappa$ & 3.89$\pm$0.04 & 4.10$\pm$0.14&4.33$\pm$0.33\\
$\alpha$& -0.14$\pm$0.04& 0.29$\pm$0.11&0.22$\pm$0.06\\
$\eta$& 0.33$\pm$0.10&4.63$\pm$1.07&1.48$\pm$0.23\\
$\theta$&-8.64$\pm$0.22&0 (fixed) &0 (fixed)\\
\hline
$\chi^{2}$/dof&0.41/33&5.29/32&0.72/23\\
\toprule
\end{tabular}
\end{center}
\end{table}

But can the CSP model describe the scaling behaviour of the charged hadron $p_{\rm T}$ spectra at the same time? The answer relies on the nonlinear trend with which the scaling parameter $K$ increases with centralities. In order to determine this trend, we fit the $K$ values for pions, kaons and protons at different centralities with a function $K = a\langle N_{\rm{part}}\rangle^{b}$, where $a$ and $b$ are free parameters, and $b$ characterizes the rate at which $\textrm{ln} K$ increases with $\textrm{ln} \langle N_{\rm{part}}\rangle$.  $a$, $b$ and the reduced $\chi^{2}$s are tabulated in table \ref{tab:a_b_parameters}. The values of $b$ for pions and kaons are consistent within uncertainties, while they are smaller than that for protons. This could also be explained by the CSP model. As described in section \ref{sec:scaling_behaviour}, the scaling parameter $K$ is proportional to $\langle p_{\rm{T}}\rangle$. Thus the ratio between the values of $K$ at non-central (5-10$\%$, 10-20$\%$, 20-40$\%$, 40-60$\%$ and 60-80$\%$) and central (0-5$\%$) collisions should be equal to the ratio between the values of $\langle p_{\rm{T}}\rangle$ at non-central and central collisions. In the CSP model, the mean $p_{\rm T}$ value is determined as
\begin{eqnarray}
\langle p_{\rm{T}}\rangle=\frac{\int_{0}^{1/p^{2}_{\rm{T}}}\int_{0}^{\infty}W(x)g(x,p_{\rm{T}})p_{\rm{T}}^{2}dxdp_{\rm{T}}}{\int_{0}^{1/p^{2}_{\rm{T}}}\int_{0}^{\infty}W(x)g(x,p_{\rm{T}})p_{\rm{T}}dxdp_{\rm{T}}}.
\label{eq:p_T_mean_CSP}
\end{eqnarray}
With the substitution of  $W(x)$ in equation (\ref{eq:gamma_function}) and $g(x,p_{\rm{T}})$ in equation (\ref{eq:frag_function}) into the above equation, we get
\begin{eqnarray}
\langle p_{\rm{T}}\rangle=\frac{\sqrt{\zeta}(\alpha+2)_{2}F_{1}(\alpha+3,-\theta;\alpha+\eta+4;-1)\Gamma(\kappa-3/2)}{(\alpha+\eta+3)_{2}F_{1}(\alpha+2,-\theta;\alpha+\eta+3;-1)\Gamma(\kappa-1)},
\label{eq:p_T_mean_r}
\end{eqnarray}
where $_{2}F_{1}$ is the hypergeometric function. Since the cluster's fragmentation function $g(x, p_{\rm T})$ is universal for a species of particles at different centralities, the parameters $\alpha$, $\eta$ and $\theta$ are the same for pions (kaons or protons) at these centralities. Thus the ratio between the values of $\langle p_{\rm{T}}\rangle$ at non-central and central collisions depends on $\zeta$ and $\kappa$. In order to determine these two parameters at non-central collisions, we fit the pion, kaon and proton spectra in the regions with $p_{\rm T} \leq$ 2.75, 3.10 and 2.35 GeV/c at these centralities to equation (\ref{eq:CPS_formula}) with $\alpha$, $\eta$ and $\theta$ fixed to their center values in table \ref{tab:CSP_fit_parameters}. $\zeta$ and $\kappa$ returned from the fits are tabulated in table \ref{tab:id_particles_gamma_kappa_parameters}. The uncertainties quoted are determined by adding the errors returned by the fits and the errors originating from the variation of the $\alpha$, $\eta$ and $\theta$ values in turn by $\pm1\sigma$ in the fits in quadrature. With these $\zeta$ and $\kappa$ values, we can calculate the ratios between the values of $\langle p_{\rm T}\rangle$ at non-central and central collisions. They are listed in table \ref{tab:pt_z_five_centralities}. Comparing these ratios with the scaling parameters $K$ at non-central collisions in table \ref{tab:id_particles_a_k_parameters} \footnote{After applying the cut $p_{\rm T} \leq 2.75$ (3.10) GeV/c to the pion (kaon) spectrum at the 60-80$\%$ centrality, the parameters $K$ and $A$ for pions (kaons) at this centrality are determined to be 0.90$\pm$0.01 and 1.27$\pm$0.13 (0.91$\pm$0.01 and 1.42$\pm$0.16).}, we find they are consistent within uncertainties. Therefore, the CSP model can also simultaneously explain the scaling behaviour of the pion, kaon and proton spectra in a quantitative way.

\begin{table}[H]
\caption{\label{tab:a_b_parameters} $a$ and $b$ returned by the fits on the scaling parameters $K$ of the pion, kaon and proton (pion and proton) spectra in Pb-Pb (Au-Au) collisions at different centralities. The uncertainties are due to the errors of $K$. The last column shows the reduced $\chi^{2}$s of the fits. } 
\begin{center}
\begin{tabular}{@{}cccc}
\toprule
\textrm{\ }&
\textrm{$a$}&
\textrm{$b$}&
\textrm{$\chi^{2}$/dof}
\\
\hline
Pions(Pb-Pb)&0.85$\pm$0.03& 0.028$\pm$0.008&7.35/4\\
Kaons(Pb-Pb)&0.83$\pm$0.03& 0.031$\pm$0.007&2.24/4\\
Protons(Pb-Pb) &0.562$\pm$0.029 &0.099$\pm$0.010&2.73/4\\
Pions(Au-Au)&0.962$\pm$0.007& 0.004$\pm$0.002&0.07/3\\
Protons(Au-Au) &0.856$\pm$0.027 &0.025$\pm$0.007&0.72/3\\
\toprule
\end{tabular}
\end{center}
\end{table}

\begin{table}[H]
  \caption{\label{tab:id_particles_gamma_kappa_parameters}$\zeta$ and $\kappa$ of the CSP fits on the pion, kaon and proton spectra at non-central collisions. The uncertainties quoted are determined by adding the errors returned by the fits and the errors originating from the variation of the $\alpha$, $\eta$ and $\theta$ values in turn by $\pm1\sigma$ in the fits in quadrature. The last column shows the reduced $\chi^{2}$s for the fits.}
\begin{center}
\begin{tabular}{@{}cccccc}
\toprule
 \textrm{\ }&
\textrm{Centrality}&
\textrm{$\zeta$}&
\textrm{$\kappa$}&
$\chi^{2}$/dof\\
\hline
\textrm{\ }&5-10$\%$&7.90$\pm$0.99&3.79$\pm$0.11&0.71/36\\
\textrm{\ }&10-20$\%$&7.66$\pm$1.01&3.72$\pm$0.12&1.06/36\\
\textrm{Pions}&20-40$\%$&6.85$\pm$0.80&3.54$\pm$0.09&0.93/36\\
\textrm{\ }&40-60$\%$&5.17$\pm$0.60&3.22$\pm$0.07&1.80/36\\
\textrm{\ }&60-80$\%$&3.78$\pm$0.42&2.95$\pm$0.05&3.84/36\\
\hline
\textrm{\ }&5-10$\%$&20.91$\pm$8.79&3.98$\pm$0.31&7.68/34\\
\textrm{\ }&10-20$\%$&19.17$\pm$8.40&3.81$\pm$0.28&5.56/34\\
\textrm{Kaons}&20-40$\%$&16.39$\pm$7.13&3.57$\pm$0.25&7.77/34\\
\textrm{\ }&40-60$\%$&11.83$\pm$5.24&3.19$\pm$0.21&6.52/34\\
\textrm{\ }&60-80$\%$&8.46$\pm$3.50&2.89$\pm$0.15&11.60/34\\
\hline
\textrm{\ }&5-10$\%$&21.63$\pm$8.29&4.35$\pm$0.83&1.26/25\\
\textrm{\ }&10-20$\%$&19.25$\pm$6.38&4.07$\pm$0.62&2.01/25\\
\textrm{Protons}&20-40$\%$&17.78$\pm$3.70&4.12$\pm$0.37&2.40/25\\
\textrm{\ }&40-60$\%$&10.79$\pm$2.24&3.39$\pm$0.22&2.91/25\\
\textrm{\ }&60-80$\%$&6.75$\pm$1.22&3.02$\pm$0.14&3.37/25\\
\toprule
\end{tabular}
\end{center}
\end{table}

\begin{table}[H]
  \caption{\label{tab:pt_z_five_centralities} The ratios between the values of $\langle p_{\rm T}\rangle$ at non-central and central collisions for pions, kaons and protons. The uncertainties quoted are due to the errors of $\zeta$ and $\kappa$.} 
\begin{center}
\begin{tabular}{@{}cccc}
\toprule
\textrm{Centrality}&
\textrm{Pions}&
\textrm{Kaons}&
\textrm{Protons}
\\
\hline
5-10$\%$&1.00$\pm$0.08&1.00$\pm$0.26&0.99$\pm$0.27\\
10-20$\%$&1.00$\pm$0.08&1.00$\pm$0.27&0.99$\pm$0.24\\
20-40$\%$&0.99$\pm$0.07&0.98$\pm$0.26&0.94$\pm$0.16\\
40-60$\%$&0.95$\pm$0.07&0.93$\pm$0.25&0.88$\pm$0.14\\
60-80$\%$&0.90$\pm$0.06&0.88$\pm$0.23&0.78$\pm$0.12\\
\toprule
\end{tabular}
\end{center}
\end{table}

Finally, we would like to see whether the centrality dependence of the scaling parameter $K$ for pions and protons in Pb-Pb collisions is the same as that in Au-Au collisions. In reference \cite{pion_spectrum}, it presented the centrality dependence of the scaling parameter $K$ for neutral pions in Au-Au collisions at 200 GeV.  In reference \cite{proton_antiproton_spectra}, it showed the centrality dependence of the scaling parameter $K$ for protons and antiprotons in Au-Au collisions at 200 GeV separately. So far, there is no information about the centrality dependence of the scaling parameter $K$ for charged pions ($\pi^{+}+\pi^{-}$) and protons ($p+\bar{p}$). Thus, in this work, in order to do the comparison, we utilize the data in reference \cite{Au_Au_collisions} and follow the procedure in section \ref{sec:method} to search for the scaling parameters $K$ for charged pions and protons at non-central collisions (10-20$\%$, 20-40$\%$, 40-60$\%$ and 60-80$\%$). The scaling parameter $K$ at central collisions (0-12$\%$) has been set to be 1. With the QF method in section \ref{sec:scaling_behaviour}, the scaling parameters $K$ for charged pions (protons) at 10-20$\%$, 20-40$\%$, 40-60$\%$ and 60-80$\%$ centralities in Au-Au collisions are 0.980$\pm$0.023,  0.979$\pm$0.015, 0.975$\pm$0.014 and  0.972$\pm$0.014 (0.983$\pm$0.016,  0.974$\pm$0.014, 0.944$\pm$0.009 and 0.930$\pm$0.015). Now we fit $K = a\langle N_{\rm{part}}\rangle^{b}$ to these $K$ values.  $a$, $b$ and the reduced $\chi^{2}$s of the fits are listed in the last two rows of table \ref{tab:a_b_parameters}. From the table, we see that for both pions and protons, the $b$ value in Pb-Pb collisions is larger than that in Au-Au collisions, which means that $\textrm{ln} K$ increases with $\textrm{ln} \langle N_{\rm{part}}\rangle$ faster in Pb-Pb collisions than in Au-Au collisions.

\section{Conclusions}\label{sec:conclusion}
In this paper, we have shown that in the low $p_{\rm T}$ region ($p_{\rm T} \leq$ 2.75, 3.10 and 2.35 GeV/c) the pion, kaon and proton spectra at different centralities in Pb-Pb collisions at 2.76 TeV exhibit a scaling behaviour. This scaling behaviour is presented when the spectra are expressed in the suitable variable, $z=p_{\rm{T}}/K$. The scaling parameter $K$ is determined by the QF method and its logarithm value is proportional to $\textrm{ln}\langle N_{\rm{part}}\rangle$. The proportionality constants for pions and kaons are consistent within uncertainties, while they are smaller than that for protons. In the high $p_{\rm T}$ region, due to the suppression of the spectra, a violation of the proposed scaling is observed going from central to peripheral collisions. The more peripheral the collisions are, the more clearly violated the proposed scaling becomes. We argue that  the pions, kaons and protons are produced by the fragmentation of clusters which are formed by strings overlapping and the fragmentation functions are different for different hadrons. The scaling behaviour of the pion, kaon and proton spectra in the low $p_{\rm T}$ region can be explained by the colour string percolation model in a qualitative way at the same time.

\section*{Acknowledgements}
We would like to thank H. Zheng and C.B. Yang for valuable discussions. This work was supported by the Fundamental Research Funds for the Central Universities of China under Grant Nos. GK201502006 and GK201803013, by the Scientific Research Foundation for the Returned Overseas Chinese Scholars, State Education Ministry, by Natural Science Basic Research Plan in Shaanxi Province of China (program No. 2017JM1040) and by the National Natural Science Foundation of China under Grant Nos. 11447024 and 11505108.


\end{document}